\documentstyle{article}

\def\be{\begin{eqnarray}}
\def\ee{\end{eqnarray}}
\def\nn{\nonumber}

\begin{document}

\phantom. \hfill ITEP-20/00 \\
\phantom. \hfill hepth/0005053

\bigskip

\centerline{\Large{Bogolubov's Recursion and}}
\centerline{\Large{ Integrability of Effective Actions}}

\bigskip

\centerline{A.Gerasimov, A.Morozov and K.Selivanov}

\centerline{\it 117259, ITEP, Moscow, Russia}

\bigskip

\centerline{ABSTRACT}

\bigskip

The Hopf algebra of Feynman diagrams, analyzed by
A.Connes and D.Kreimer,
is considered from the perspective of the theory of effective
actions and generalized $\tau$-functions, which
describes the action of diffeomorphism and shift groups
in the moduli space of coupling constants.
These considerations provide additional evidence of the hidden group
(integrable) structure behind
the standard formalism of quantum field theory.

\tableofcontents

\section{Introduction}
\setcounter{equation}{0}

Exponentiated effective action (partition function, statistical sum)
is defined as a functional of the coupling
constants $T$ and background fields $\varphi$ (the ``vacuum
configuration''), resulting from functional integration
over quantum fields $\phi$:

\be
{\cal Z}\{T|\varphi\} = \int
\exp \left(-S_T(\varphi + \phi)\right) {\cal D}\phi
\label{effa}
\ee
When all possible coupling constants $T$ are taken into account
(i.e. the theory is maximally deformed), ${\cal Z}\{T\}$ becomes a
generating function of all the correlation functions in entire
family of models.
Such ${\cal Z}\{T\}$ possesses a hidden group-theoretical
structure and -- as a manifestation of this --
satisfies bilinear (Hirota-like) and differential (Laplace-like)
equations, i.e. belongs to the class of generalized
$\tau$-functions.
One can consider ${\cal Z}\{T\}$
as a function (section) on the moduli space ${\cal M}$ of theories
(parametrized by the coupling constants $T$).
There are two important groups, acting transitively on ${\cal M}$:
the abelian group ${\it Shift}{\cal M}$ of shifts along ${\cal M}$
and non-abelian group ${\it Diff}{\cal M}$ of diffeomorphisms
of ${\cal M}$. They act on partition functions in the same way:

\be
{\cal Z}\{T\} \longrightarrow {\cal Z}\{T + V(T)\},
\ee
but the composition rules are different:

\be
{\cal Z}\{T\} \longrightarrow {\cal Z}\{T + V_1(T) + V_2(T)\}
\ee
for ${\it Shift}{\cal M}$  and

\be
{\cal Z}\{T\} \longrightarrow {\cal Z}\{T + V_1(T) + V_2(T + V_1(T))\}
\ee
for ${\it Diff}{\cal M}$.
In other words, the infinitesimal action of both groups is
decribed by vector fields $\hat V\{T\} = V(T)\partial/\partial T$,
but the global action is by exponentiated vector field,
$\exp \hat V$ for  ${\it Diff}{\cal M}$ and by the normal-ordered
exponent $:\exp \hat V:$ for abelian ${\it Shift}{\cal M}$.
A map between ${\it Diff}{\cal M}$ and ${\it Shift}{\cal M}$
is provided by the relation

\be
e^{\hat V} = \ :e^{\hat{\tilde V}}:
\label{difvershift}
\ee
Looking from this perspective, one associates with every
particular theory a group element $g_T$
with two basic properties

\be
g_{T_1} = g_{T_{12}} g_{T_2}, \ \ \ T_1 = T_{12} \circ T_2
\label{mult}
\ee
and

\be
\Delta g_T = g_T \otimes g_T,
\label{comult}
\ee
and represents ${\cal Z}\{T\}$ as a matrix element (generalized
zonal-function or $\tau$-function):
${\cal Z}\{T\} = \langle 1 | g_T | 0 \rangle$  between a Gaussian theory,
labeled by $| 0 \rangle$, and some other state $\langle 1|$,
depending on particular realization of $g_T$.
The transitive action of the group basically puts all
the points in the moduli space on equal footing (in particular
the Gaussian point is not distinguished among the others),
and this can explain the surprising power of the free-field formalism in
quantum field theory.
The composition rule
$T_1 = T_{12} \circ T_2$
depends on which of the two groups,
${\it Shift}{\cal M}$  or ${\it Diff}{\cal M}$,
we want $g_T$ to belong to. In the case of abelian
${\it Shift}{\cal M}$  it is just an addition: $T_1 = T_{12} + T_2$
(if the space of coupling constants is big
enough and appropriate choice of coordinates in the moduli
space is made).
A more interesting non-abelian diffeomorphism group,
and especially the stability subgroup ${\it Diff}_\emptyset{\cal M}$
of the Gaussian model in ${\cal M}$,
is relevant for description of one-parametric
renormalization-group flows along ${\cal M}$.

This old set of ideas \cite{GLM, GMMOS, UFN2, Hirota, gentau}
is supported by evidence in matrix models \cite{UFN3},
Seiberg-Witten theory \cite{GKMMM,RG,Edi99} and AdS/CFT
correspondence \cite{Mor98,Ver}\footnote{
The AdS/CFT-correspondence \cite{AdS}
claims that certain Yang-Mills partition functions
are represented by the boundary dependence of the
bulk actions of certain classical gravities. Among other things,
this implies that they satisfy bilinear Hamilton-Jacobi equations,
which should be nothing but an avatar of bilinear Hirota
and Laplace-like equations for the effective actions.
}.
The purpose of this paper is to claim that additional evidence
is provided by the recent studies of A.Connes and D.Kreimer
\cite{CK}, who actually define the action of the operators $g_T$
in the space of graphs (Feynman diagrams).
In what follows we basically repeat their reasoning,
making use of convenient quantum models
and separating the algebraic structures, relevant for
Hirota-like equations and Bogolubov's $R$-operation,
from peculiarities of particular models and prejudices
of conventional local field theory.
Some graph-theory and combinatorial routine is omitted.

\section{The Basics of Bilinear Relations}
\setcounter{equation}{0}

The basic model for generic studies of integrable
structures in quantum field theory is the one of arbitrarily many
scalars $\phi^i$ in $0+0$ dimensions:

\be
{\cal Z}\{T\} =
\int   e^{ V_T(\phi)} \prod_i d\phi^i,
\label{genthe}
\ee
\be
V_T(\phi) = \sum_n \frac{1}{n!}\left( \sum_{i_1,\ldots,i_n}
T^{(n)}_{i_1\ldots i_n}\phi^{i_1}\ldots\phi^{i_n}\right)
\label{genpot}
\ee

Partition function
${\cal Z}\{T\}$ can be represented as a sum over all possible
graphs without external legs (vacuum Feynman diagrams).
One can exclude disconnected graphs by switching to
$\log{\cal Z}\{T\}$ -- in exchange one gets additional
$1/n$ coefficients. These can be eliminated by consideration
of correlation functions: graphs with external legs.
This is the usual routine of diagram technique \cite{AGD}.

At this point it can make sense to comment on the
choice of the model (\ref{genthe}). Among its
particular reductions (truncations) are:
the single-scalar model,

\be
{\cal Z}_{U(1)}\{T\} =
\int \exp \left(\sum_{n=0}^\infty \frac{T_n}{n!}\phi^n\right) d\phi;
\label{1sthe}
\ee
the $N\times N$ matrix model with $N^2$ scalars, assembled into
a matrix $\phi_{ab}$,

\be
{\cal Z}_{U(N)}\{T\} =
\int  \exp
\left(\sum_{n=0}^\infty  \frac{T_n}{n!}
Tr \phi^n\right) \prod_{a,b=1}^N d\phi_{ab};
\label{mamthe}
\ee
the scalar field in $d$ space-time dimensions, where indices $i$
become continuous and $T^{(2)}_{ij}$ is taken to be Laplace operator;
etc.
(In the last example, if the space-time is non-compact, it is
unavoidable to introduce the background fields $\varphi$,
like it is done in (\ref{effa}), to label the boundary conditions and/or
the asymptotics at infinities.)
The most essential difference of all these popular models
from the universal one in (\ref{genthe}) is that the
vertices in (\ref{genthe}) are of the most general form,
e.g. in $d$-dimensional theory the $\phi^n$ coupling should
allow {\it any} dependence of the coupling ``constants'' on all the
$n$ $d$-momenta (while in local field models they are indeed
constants or at best polynomia in momenta). In such a large
moduli space one can distinguish between any two Feynman diagrams,
looking at their $T$-dependencies (while for the model (\ref{1sthe})
all the diagrams with the same number of propagators, $l$, and
vertices of valences $k$, $v_k$, give rise to the same expression,
$T_2^{-l}\prod_k T_k^{v_k}$;
and switching to the matrix model (\ref{mamthe}) introduces
nothing more than
extra factor $N^{-\chi}$, depending on Euler characteristics $\chi$
of the corresponding fat graph,
which is still not enough to distinguish between
any two diagrams). Last, but not the least, such a large
moduli space is preserved by renormalization group flow:
effective actions at any stage of the flow remains in the class
(\ref{genthe}); also renormalizability is not a restriction on
the form of the theory (once it is somehow regularized).

A further extension of (\ref{genthe}), playing the same
role of the universal model
for {\it fat} graphs as  the model (\ref{genthe})
plays for the ordinary ones, is provided by the matrix model
with 2-index fields $\phi^{ij}$ and the action
\be
V_T(\phi) = \sum_n \frac{1}{n!}\left( \sum_{i_1,\ldots,i_n}
T^{(n)}_{i_1\ldots i_n}\phi^{i_1i_2}\phi^{i_2i_3}
\ldots\phi^{i_ni_1}\right)
\label{genmathe}
\ee
Its principal difference from (\ref{genthe}) is that the
couplings $T^{(n)}_{i_1\ldots i_n}$ are no longer symmetric
under permutations of indices $i_1,\ldots,i_n$, only cyclic
symmetric.
This model remains beyond our consideration in the present
paper.

The model (\ref{genthe}) can be also regarded in a different way.
Given any particular quantum theory one can switch to its
$GL(N)$ or $GL(\infty)$ extension, just adding a vector index $i\in I$
to all the fields of the theory and ascribing the  relevant tensor
structure to all the coupling constants. For example, the $d$-dimensional
$\phi^3$ theory can be substituted by

\be
\int \prod_i D\phi^i(x) \exp \int_{d^dx} \left(
T_{ij}^{(2)}\left((\nabla\phi^i)(\nabla\phi^j) - m^2\phi^i\phi^j\right)
- g_3 T_{ijk}^{(3)}\phi^i\phi^j\phi^k
\right)
\ee
without requiring that $T$-variables are $x$-dependent. Then the
partition function is $GL(\infty)$-invariant and can be
expanded in a series over the basic $GL(\infty)$-invariant functions,
provided by (\ref{genthe}), however the expansion
coefficients are now sophisticated functions not only of
graphs, but also of many other parameters, including
external momenta and the spins of particles. Still, some
basic properties can be seen  at the level of graph theory
alone -- and this is the subject of our futher considerations.

We now return to the main line of discussion. The issue
of our interest is the group (integrable) structure, hidden
in partition functions ${\cal Z}\{T\}$.
This structure survives various reductions, e.g. the one
to the matrix model (\ref{mamthe}), see \cite{mamo},
but many aspects are much more transparent in analysis of
the universal model (\ref{genthe}).
Of course, the relation
\be
V_{T+T'} = V_T + V_{T'}
\label{add}
\ee
for the potential under the integral
does {\it not} imply that the average ${\cal Z}\{T\}$ is a character
of ${\it Shift}{\cal M}$ group,
${\cal Z}\{T+T'\} \neq {\cal Z}\{T\}{\cal Z}\{T'\}$, the true relation
is between the group elements,
$g_{T+T'} = g_T g_{T'}$, and

\be
{\cal Z}\{T+T'\} =\ \langle 1 |g_{T+T'}| 0 \rangle\ =
\sum_{states} \langle 1 |g_{T}| states \rangle
\langle states |g_{T'}| 0 \rangle
\ee
At the r.h.s. stands a non-trivial operator, acting on
$T$ and $T'$ in the product ${\cal Z}\{T\}{\cal Z}\{T'\}$.

The question is, what are the relevant realizations of the Hilbert
space of $|states\rangle$ and of the operators $g_T$ acting in it.

The simplest realization is implied by the functional
integral and makes special use of the source-dependence of
${\cal Z}\{T\}$. Namely, the $T^{(1)}$-terms in (\ref{genpot})
can be identified with the sources for the fields $\phi^i$:
$\sum_j T^{(1)}_j\phi^j = i\sum_j J_j\phi^j$ and, as usual
in the derivation of Hirota equations, one can use sources to
construct a delta-function projector:

\be
\int \ e^{-i(\phi^j-\phi'^j)J'_j} \prod_j dJ'_j  \sim
\prod_j \delta (\phi^j - \phi'^j),
\ee
so that

\be
\int {\cal Z}_{J-J'}\{T\} {\cal Z}_{J'}\{T'\}  dJ' = \nn \\
= \int dJ' \left(\int d\phi e^{V_T(\phi)}e^{i(J-J')\phi}
\int d\phi' e^{V_{T'}(\phi)}e^{i(J-J')\phi'}\right)
= {\cal Z}_J\{T+T'\}
\ee
(we explicitly labeled the $J$- ($T^{(1)}$-) dependence
of the action, suppressed all the indices $i$ and made use of
the addition formula (\ref{add})).

The simplest example arises if all $T^{(n)} = 0$ for $n>2$.
Then

\be
{\cal Z}_{Gauss}\{T\} \sim \frac{1}{\sqrt{\det T^{(2)}}}
\exp \left(- \frac{1}{4}Tr\ T^{(1)}\frac{1}{T^{(2)}}T^{(1)}\right),
\ee
and the bilinear relation

\be
{\cal Z}_{Gauss}\{T+T'\} =
\int dT^{(1)} dT'^{(1)}\delta (T^{(1)} - T'^{(1)})
{\cal Z}_{Gauss}\{T\}{\cal Z}_{Gauss}\{T'\}
\ee
is just the completeness formula for Guassian propagators,
which in the single-scalar case is widely-known in the form

\be
\int dx_{2} \frac{e^{-x_{12}^2/4t_{12}}}{\sqrt{t_{12}}}\cdot
\frac{e^{-x_{23}^2/4t_{23}}}{\sqrt{t_{23}}} \sim
\frac{e^{-x_{13}^2/4t_{13}}}{\sqrt{t_{13}}}
\ee
(in our case $x_{12} = x_1 - x_2 = T^{(1)}$, $x_{23} = T'^{(1)}$,
$t_{12} = T^{(2)}$, $t_{23} = T'^{(2)}$).

In a similar way one can obtain bilinear integral
``summation formulae'' for the Eiry functions (if all $T^{(n)} = 0$
for $n>3$) etc.

Instead of using the source-dependencies, one can exploit
those on other coupling constants. All these dependencies
are interrelated: in the universal model (\ref{genthe})

\be
\frac{\partial{\cal Z}}{\partial T^{(n)}_{i_1\ldots i_n}} =
\frac{\partial^n{\cal Z}}{\partial T^{(1)}_{i_1}\ldots
\partial T^{(1)}_{i_n}},
\label{tnt1}
\ee
in its various reductions one has more sophisticated Ward
identities, like Virasoro and W-constraints in matrix models
\cite{DVV,mamo,Vir}. Also, Legendre transform relates source- and
background-field dependencies of generic ${\cal Z}\{T|\varphi\}$
in (\ref{effa}). For more discussion of interplay between
the source-, coupling-constants and background-fields dependencies
see \cite{UFN3}, especially the example of generalized Kontsevich
model \cite{GKM}. The Ward identities like (\ref{tnt1}) play
important role in building explicit maps like (\ref{difvershift})
between the ${\it Shift}{\cal M}$ and ${\it Diff}{\cal M}$ groups.

Drawback of such functional-integral approaches to bilinear
identities is that they do not immediately provide representations in
terms of conventional (perturbative) correlation functions:
at least some operator, like the source term $\sum_j J_i\phi^i$,
should be exponentiated, i.e. one needs to consider a global
deformation and the entire family of theories, not just
infinitesimal vicinities of the given models $g_T$ and $g_{T'}$.
Though there is nothing bad about this from the general
perspective of string theory, such representations are not
the best ones for the search of bilinear relations in
conventional quantum field theories, where isolated points
in the moduli spaces (i.e. isolated particular models)
are usually analyzed.
One possibility to obtain representations in terms of
the ordinary Green functions is to use the Vermat-module
realizations of $|states\rangle$ \cite{Hirota}.
Another -- not unrelated -- possibility is
exploited by A.Connes and D.Kreimer (CK) in \cite{CK}:
it is to look at the contributions of particular graphs
(Feynman diagrams).

\section{Hilbert Space of Graphs and Operators $g_T$ \label{SPAG}}
\setcounter{equation}{0}

Operators, acting in the Hilbert space of Feynman diagrams
naturally appear if the Gaussian measure is explicitly
extracted from $\ \exp V_T$ (as a starting point for
perturbation expansion)
and if the theory (\ref{genthe}) is further
``complexified'':

\be
{\cal Z}\{\tilde T,T\} =
\int  \left\{\prod_{i,\tilde i} d\phi^{i\tilde i}
\exp \left(-\frac{1}{2}\sum_{i,j,\tilde i,\tilde j}
G_{ij}\tilde G_{\tilde i\tilde j}
\phi^{i\tilde i}\phi^{j\tilde j}\right)
\right\}
e^{ V_{\tilde T,T}(\phi)},
\label{genbithe}
\ee
\be
V_{\tilde T,T}(\phi) = \sum_n \frac{1}{n!} \left(
\sum_{i_1,\ldots,i_n;\tilde i_1,\ldots,\tilde i_n}
\tilde T^{(n)}_{\tilde i_1,\ldots,\tilde i_n}T^{(n)}_{i_1\ldots i_n}
\phi^{\tilde i_1 i_1}\ldots\phi^{\tilde i_n i_n}\right)
\label{genbipot}
\ee
The fields  $\phi^{i\tilde i}$ are now labeled by a pair of
indices, taking values in the sets $I$ and $\tilde I$,
$i\in I$, $\tilde i \in \tilde I$,
and coupling constants are factorized (assumed to be the
``squared modules'' of ``holomorphic'' $T$'s). In variance
with ordinary complexification, we do not assume that the
sets $I$ and $\tilde I$ are the same. In particular, we can
return to the original model (\ref{genthe}) by asking $\tilde I$
to consist of a single element and putting all
$\tilde T^{(n)}_{1\ldots 1} = 1$ and $\tilde G_{11}=1$
(below,when necessary, we just write $\tilde T = 1$,
implying that $\tilde I = \{1\}$, and
${\cal Z}\{T\} = {\cal Z}\{\tilde T =1,T\}$).

Expanding $\exp V_{\tilde T,T}(\phi)$ in (\ref{genbithe})
into formal series and
applying the Wick theorem to Gaussian integrals,
one obtains the expansion over vacuum Feynman diagrams $\Gamma^{(0)}$,
which has specific structure, called ``holomorphic factorization'':

\be
{\cal Z}\{\tilde T, T\} = \sum_{\Gamma^{(0)}}
\frac{
{\cal Z}_\Gamma\{\tilde T\}  {\cal Z}_\Gamma\{T\} }{S_\Gamma}
\label{holfac}
\ee
Here $S_\Gamma$ is the ``symmetry factor'' of the graph $\Gamma$,
for connected graph it is the order
of the discrete group which permutes links,
while keeping their ends fixed.
For vacuum diagrams (graphs without external legs) $S_\Gamma$
contains an additional factor ${\it Vert}(\Gamma)$ -- the number
of vertices in the graph \cite{AGD}. For disconnected graphs
$S(\prod_i\Gamma_i^{n_i}) = \prod_i n_i! S_{\Gamma_i}^{n_i}$
Expression $Z_\Gamma\{T\}/S_\Gamma$
for the Feynman diagram $\Gamma$ is a convolution of vertices
and propagators, divided by $S_\Gamma$.
In particular, $Z_\Gamma\{T=1\} = 1$, and
for the model (\ref{genthe}) eq.(\ref{holfac}) gives:
\be
{\cal Z}\{T\} =
{\cal Z}\{\tilde T=1,T\} = \sum_{\Gamma^{(0)}} \frac{
{\cal Z}_\Gamma\{T\}  }{S_\Gamma}
\label{holfac1}
\ee
In Feynman diagrams $G_{ij}$ plays the role of inverse propagator.
In what follows we lower and raise indices with the help of $G_{ij}$
and its inverse $G^{ij}$. In particular the switch from Green functions
to the amputated correlators is nothing but lowering of indices on
external legs. Coupling constants are defined to have lower indices.

Introduce now the Hilbert space ${\cal H}^{(0)}$ of
all possible graphs (with vertices of any valence,
connected or disconnected) with no external legs (vacuum
Feynman diagrams), i.e. with every graph $\Gamma^{(0)}$
we associate a state $| \Gamma^{(0)} \rangle$.
The scalar product is

\be
\langle\Gamma^{(0)} | \Gamma'^{(0)} \rangle \ =
S_\Gamma \delta_{\Gamma, \Gamma'}
\ee
Let us further define the coherent-like states

\be
|T\rangle = \sum_{\Gamma^{(0)}}
\frac{{\cal Z}_\Gamma\{T\} }{S_\Gamma} |\Gamma\rangle
\label{defTstate}
\ee
They are not orthonormal, instead (\ref{holfac}) states that

\be
{\cal Z}\{\tilde T, T\} = \ \langle \tilde T | T\rangle
\ee
Thus with every particular model (a point in the moduli space
${\cal M}$) we associate a point $|T\rangle$ in the Hilbert space of
vacuum graphs, and partition function ${\cal Z}\{\tilde T,T\}$
(\ref{genbithe}) is just a scalar product of associated
states. In particular, the holomorphic
partition function ${\cal Z}\{T\}$
in (\ref{genthe}) is a scalar product of $|T\rangle\ $
with a special state $\langle \tilde T=1|$,
where $\tilde T$ has single-valued indices and all $\tilde T=1$.

Another special state is the Gaussian model,
$|Gauss\rangle\ = |0\rangle\ = |\emptyset\rangle$,
associated with all $T = 0$ (with any set $I$ and non-vanishing
metric $G_{ij}$). At Gaussian point the only contribution
to (\ref{defTstate}) comes from the empty graph $\Gamma = \emptyset$
(we assume the normalization ${\cal Z}\{T=0\} = 1$).
There is an even more distinguished ``trivial''
model with ${\cal Z}\{T\} =
\exp T^{(0)}$, the
transitive actions of ${\it Shift}{\cal M}$ and ${\it Diff}{\cal M}$
groups connect it to any other model, including the Gaussian one.
However, the structures which are of interest for us are explicitly
dependent on the metric $G_{ij}$, and the minimal model which
takes this into account is the Gaussian one: arbitrary diffeomorphisms
and shifts can be expanded over basic functions, provided by Gaussian
model, but not by the ``trivial'' one.

Let us now introduce an operator $g_T$, acting in the Hilbert space
of graphs, such that

\be
|T\rangle\ = g_T|Gauss\rangle\ = g_T|\emptyset\rangle
\ee
Then

\be
{\cal Z}_\Gamma\{T\} = \langle \Gamma | T \rangle =
\langle\Gamma|g_T|\emptyset\rangle
\label{gT}
\ee
In sec.\ref{reps} below we use these matrix elements to convert
functions of coupling constants $T$ into functions of graphs
$\Gamma$ and vice versa.

Eq.(\ref{gT}) does not fully specify the operator $g_T$.
However, field theory implies a natural extension
of (\ref{gT}) to all the matrix elements

\be
\langle \Gamma^{(n)}_{i_1\ldots i_n}|g_T|
\gamma^{(m)}_{j_1\ldots j_m}\rangle
\label{mael}
\ee
in the enlarged Hilbert space ${\cal H} = \oplus_n
{\cal H}^{(n)}$ of all the graphs $\Gamma^{(n)}$ with any number $n$
of external legs with indices $i\in I$ ascribed to every leg.

The space ${\cal H}^{(n)}$ naturally appears if one
considers the $n$-point correlation functions in the theory
(\ref{genbithe}). Such correlator carries extra $2n$ indices
and is decomposed into a sum over all the graphs with
$n$ external legs:

\be
{\cal Z}\{\tilde T, T\}^{\tilde i_1\ldots\tilde i_n;i_1\ldots i_n}
= \sum_{\Gamma^{(n)}} \frac{
{\cal Z}_\Gamma\{\tilde T\}^{\tilde i_1\ldots\tilde i_n}
{\cal Z}_\Gamma\{T\}^{i_1\ldots i_n}  }{S_\Gamma}
\label{holfac2}
\ee
Similarly to the case of ${\cal H}^{(0)}$
one can now define a set of states
$|\Gamma_{(n)}^{i_1\ldots i_n}\rangle$ with the scalar product

\be
\langle \Gamma_{(n)}^{i_1\ldots i_n} |
{\Gamma'_{(m)}}^{j_1\ldots j_m} \rangle\
= S_\Gamma
G^{i_1j_1}\ldots G^{i_nj_n}\delta_{\Gamma,\Gamma'}
\ee
(Since scalar product is non-vanishing only for coincident graphs,
the number of external legs are also the same, and $G^{i_kj_k}$
couples the indices ascribed to the same $k$-th leg.)
One can amputate external legs by lowering the indices with the help
of the metric $G_{ij}$.\footnote{
When one cuts a link in a graph, two new external legs are formed
at the place of a single propagator $G^{-1}$ and one glues them
back with the help of the metric $G$, or, alternatively, amputate
one leg in each pair. This can be done more symmetrically, if
$G = D^2$: then one can associate with every external leg the matrix
$D^{-1}$, instead  of the usual rule, ascribing the propagator $G^{-1}$
to non-amputated leg and unity to the amputated one.
In continuous case, when $G$ is Laplace operator,
$D$ turns into a Dirac operator.
Though the use of $D$ can make the bilinear relations below
conceptually more symmetric, we ignore this possibility in the
present text.
}

Now we introduce in ${\cal H}^{(n)}$ the state

\be
|T\rangle^{(n)} = \sum_{\Gamma^{(n)}; i_1,\ldots, i_n }
\frac{
{\cal Z}_\Gamma\{T\}^{i_1\ldots i_n} }{S_\Gamma}
|\Gamma^{(n)}_{i_1\ldots i_n}\rangle
\ee
(the indices are lowered with the help of the metric $G$)
and finally the state
\be
| T \rangle \ = \oplus_n |T\rangle^{(n)}
\ee
in entire ${\cal H}$.

In order to define the matrix elements of $g_T$ between any two
states in ${\cal H}$ we need to introduce the notion of subgraph.

\section{Subgraphs}
\setcounter{equation}{0}

There are two different notions of subgraph, relevant
for our further discussion.
Let $\Gamma^{(n)}$ be a graph (connected or disconnected, possibly
one-particle reducible) with $n$ external legs. It has vertices
of any valence (including one and two).

{\bf 1) Vertex-subgraps.}
Divide the set of vertices in
two non-in\-ter\-sec\-ting subsets and cut all the links,
connecting vertices from different sets. If $m$ legs were cut,
we decompose the original graph $\Gamma^{(n)}$ into two disconnected
graphs $\gamma_1^{m+n_1}$ and $\gamma_2^{m+n_2}$, such that
$n_1+n_2 = n$. We call them {\it vertex}-subgraphs of $\Gamma^n$ and
introduce a notation $\gamma_2 = \Gamma/\gamma_1$ (of course, also
$\gamma_1 = \Gamma/\gamma_2$).
The empty graph $\gamma^{(0)} = \emptyset$
and $\gamma^{(n)} = \Gamma^{(n)}$ are vertex-subgraphs of $\Gamma^{(n)}$.
The number of vertices $Vert(\gamma) + Vert(\Gamma/\gamma) =
Vert(\Gamma)$.

{\bf 2) Box-subgraphs.}
Pick a non-empty\footnote{
The would-be empty box is not well defined. If there are no
vertices inside the box, it still can be not empty:
contain fragments of some links.
To avoid such ambiguities we
exclude empty graphs from the set of box-subgraphs of $\Gamma$.
When necessary, their contributions will be explicitly added to
sums over the set ${\cal B}\Gamma$ of box-subgraphs.
}
subset of vertices and draw a
box or a set of non-intersecting boxes
around them. Boxes should not lie one inside another.
Each box in the set should contain at least one vertex,
and the subgraph inside the box should be connected.
The sides of the box cut some links of original graph,
in particular a link connecting two vertices from our
subset can be cut (and these two vertices can belong
to the same box or to two disconnected boxes). We call the
subgraph $\gamma$ lying in this system of boxes a
{\it box}-subgraph of $\Gamma^{(n)}$.
Its complement is no longer a box-subgraph: it can contain
just a remnant of a double-cut link with no vertices.
Instead of a complement, for a box-subgraph $\gamma^{(m)}$
one can always define a contraction
$[\Gamma^{(n)}/\gamma^{(m)}]$ obtained when each connected
component of a box, which cuts links at $k$ places is substituted
by a single valence-$k$ vertex. The resulting graph
$[\Gamma^{(n)}/\gamma^{(m)}]$ has $n$ external legs, as the
original $\Gamma^{(n)}$ and the same number of connected
components, ${\it Con}([\Gamma/\gamma]) = {\it Con}(\Gamma)$.
According to this definition, the empty graph
$\emptyset$ is not a box-subgraph
of $\Gamma$, and there is no box-subgraph $\gamma$, such that
$[\Gamma/\gamma] = \emptyset$.
The number of vertices $Vert(\gamma) + Vert([\Gamma/\gamma])
= Vert(\Gamma) + {\it Con}(\Gamma)$.

The same graph $\gamma$ can happen to be a vertex-subgraph
and a box-subgraph simultaneously, but the two sets
${\cal V}\Gamma$ and ${\cal B}\Gamma$
(of vertex- and box-subgraphs respectively) are different.
${\cal V}\Gamma$ is just a set-theory object:
the set of all subsets of the set of vertices of $\Gamma$,
in particular there are always exactly $2^{Vert(\Gamma)}$
vertex-subgraphs. As to ${\cal B}\Gamma$, this is a more
sophisticated object, essentially depending on the graph structure
(not just the set-theory one), in particular, the size
of this set depends on the valences of vertices and
on exact construction of the links.

The set ${\cal V}\Gamma$ is related to the abelian
group ${\it Shift} {\cal M}$
(and is relevant for description of bilinear identities),
while ${\cal B}\Gamma$ is related to the non-abelian
group ${\it Diff}_\emptyset{\cal M}$, generated by vector
fields on ${\cal M}$ (and is relevant for description of
Bogolubov's recursion and renormalization flows).

\subsection{Examples}

It is now instructive to consider some examples.
We mention three classes of simple graphs, useful for
various illustrations.

{\bf 1) Single-vertex graph} $\Gamma_{p;n}^{(p-2n)}$
has one valence-$p$ vertex, $n$ propagators  and
$p-2n$ external legs. $\Gamma_{p;0}^{(p)}$ is the
elementary (bare) vertex of valence $p$.

$\Gamma_{p;n}^{(p-2n)}$
has just two vertex-subgraphs: $\gamma = \emptyset$ and
$\gamma = \Gamma_{p;n}^{(p-2n)}$ itself.
The corresponding complements are $\Gamma_{p;n}^{(p-2n)}/\emptyset =
\Gamma_{p;n}^{(p-2n)}$
and $\Gamma_{p;n}^{(p-2n)}/\Gamma_{p;n}^{(p-2n)} = \emptyset$.

At the same time, there are $2^n$ box-subgraphs (of which $n+1$ are
topologically different):
$\gamma = \frac{n!}{k!(n-k)!}\times \Gamma_{p;k}^{(p-2k)}$,
$0\leq k \leq n$ (binomial coefficient ${n!}/{k!(n-k)!}$
denotes the multiplicity of the subgraph).
The corresponding  contractions
$\left[\Gamma_{p;n}/\Gamma_{p;k}\right]$ are
$\frac{n!}{k!(n-k)!}\times\Gamma_{p-2k;n-k}^{(p-2n)}$.

{\bf 2) Two-vertex graph} $\Gamma_{p,q;n}^{(p+q-2n)}$
has one valence-$p$ and one valence-$q$ vertices, $n$ propagators
{\it between} them (i.e. $0\leq n \leq p,q$) and
$p+q-2n$ external legs.

It has $4$ vertex-subgraphs,
$\gamma = \emptyset, \ \Gamma_{p;0},\ \Gamma_{q;0},\
\Gamma_{p,q;n}$, and $2^n + 2$ box-subgraphs:
$\gamma = \Gamma_{p;0},\ \Gamma_{q,0}, \
\frac{n!}{k!(n-k)!} \Gamma_{p,q;k}, \ 0\leq k \leq n$.
The corresponding $[\Gamma/\gamma] =
\Gamma_{q;0},\ \Gamma_{p;0},\
\frac{n!}{k!(n-k)!}\times\Gamma_{p+q-2k; n-k}$.

{\bf 3) Chain graph} $C_N$ has $N$ valence-two vertices,
connected chain-wise by $N-1$ propagators. It has $2$ external legs.

$C_N$ has $2^N$ vertex-subgraphs and $\beta_N$ box-subgraphs.

{\bf $N=1$.}

Vertex subgraphs:

\be
\begin{array}{cccc}
\gamma & = & \emptyset, & C_1 \\
C_1/\gamma & = & C_1, & \emptyset
\end{array}
\ee

Box-subgraphs ($\beta_1 = 1$):

\be
\begin{array}{ccc}
\gamma & = &  C_1 \\
\phantom. [C_1/\gamma] & = & C_1
\end{array}
\ee

{\bf $N=2$.}

Vertex subgraphs:

\be
\begin{array}{ccccc}
\gamma & = & \emptyset, & 2\times C_1, & C_2 \\
C_2/\gamma & = & C_2, & 2\times C_1, & \emptyset
\end{array}
\ee

Box-subgraphs ($\beta_2 = 4$):

\be
\begin{array}{ccccc}
\gamma & = &  2\times C_1, & C_2, & C_1\cdot C_1 \\
\phantom. [C_2/\gamma] & = & 2\times C_2, & C_1, & C_2
\end{array}
\ee

{\bf $N=3$.}

Vertex subgraphs:

\be
\begin{array}{cccccccc}
\gamma & = & \emptyset, & 2\times C_1, & C_1, & 2\times C_2, &
C_1\cdot C_1, & C_3 \\
C_3/\gamma & = & C_3,& 2\times C_2, & C_1\cdot C_1, &
2\times C_1, & C_1, & \emptyset
\end{array}
\ee

Box-subgraphs ($\beta_3 = 12$):

\be
\begin{array}{cccc}
\gamma & = & 3\times C_1, & 2\times C_2, \\
\phantom. [C_3/\gamma] & = & 3\times C_3, & 2\times C_2,
\end{array} \nn \\
\begin{array}{ccccc}
C_3, & 2\times (C_1\cdot C_1),
& C_1\cdot C_1, & 2\times (C_1\cdot C_2), & C_1\cdot C_1\cdot C_1 \\
C_1, & 2\times C_2, &
C_3, & 2\times C_2, & C_3
\end{array}
\ee

Every box-subgraph of $C_N$ is located in $s$ non-intersecting
boxes, with $k$-th box beginning at link $i_k$
and ending at link $j_k$. The total number

\be
\beta_N = \sum_{s=1}^N \beta(N;s) = \nn \\ =
\sum_{s=1}^N\left(\sum_{0\leq i_1 < j_1 \leq i_2 < j_2 \leq \ldots
\leq i_s < j_s \leq N} 1 \right) = \mu_{N-1}\mu_N
\ee
The number of chain graphs with exactly $s$ connected components is

\be
\beta(N;s) = \frac{(N+s)!}{(2s)!(N-s)!}
\ee
The $\mu_N$ are Fibonacchi-like numbers, satisfying recurrent
relations:

\be
\mu_{2k} = 4\mu_{2k-1} - \mu_{2k-3},\ \ \
\mu_{2k+1} = 3\mu_{2k-1} - \mu_{2k-3}
\ee
and initial conditions $\mu_0 = \mu_1 = 1$. Consequently

\be
(\mu_0,\mu_1,\mu_2,\ldots ) =
(1,1,4,3,11,8,29,21,76,55,\ldots\ )
\ee
and

\be
(\beta_1,\beta_2,\ldots ) = (1,4,12,33,88,232,609,1596,\ldots\ )
\ee

\section{Vertex-subgraphs,
action of $g_T$ in ${\cal H}$ and bilinear relations}
\setcounter{equation}{0}

We are now ready to define the matrix elements of $g_T$.
They are different from zero only for $\gamma$ which is
a vertex-subgraph of $\Gamma$ (consequently, $g_T$ is triangular
and can not
be Hermitean operator -- this is natural, since it
is an element of a group, not algebra,-- moreover,
triangularity implies that $g_T^\dagger \neq g_{-T}$).
For simplicity we first assume that
no external legs of $\Gamma^{(n)}$
were cut to make the subgraph $\gamma^{(m)}$. Then

\be
\langle \Gamma_{(n)}^{i_1\ldots i_n} |
g_T | \gamma_{(m)}^{j_1\ldots j_m} \rangle\ =
Z_{\Gamma/\gamma}^{i_1\ldots i_nj_1\ldots j_m}\{T\}
\label{me}
\ee
In other words, the matrix element is given by
the expression for Feynman diagram $\Gamma/\gamma$
in the theory $g_T$ without the usual symmetry factor
$1/S_{\Gamma/\gamma}$.
This means that every link in $\Gamma/\gamma$
carries a propagator $G^{ij}$ and every vertex of valence $k$
in $\Gamma/\gamma$ contributes $T^{(k)}_{i_1\ldots i_k}$.
The indices are contracted and summed over. Since
$\Gamma^{(n)}/\gamma^{(m)}$ has $n$ original and $m$
new-formed external legs,
the whole matrix element has $n+m$ free indices.
If some $p$ of external legs of $\gamma$ coincide with external
legs of $\Gamma$, the corresponding indices appear as
$\delta^i_j$ (or $G^{ij}$) factors.

\be
\langle \Gamma_{(n)}^{i_1\ldots i_{n-p}l_1\ldots l_{p}} |
g_T | {\gamma_{(m)}^{j_1\ldots j_{m-p}}}_{k_1\ldots k_{p}}\rangle\ =
Z_{\Gamma/\gamma}^{i_1\ldots i_{n-p}j_1\ldots j_{m-p}}\{T\}
\delta^{l_1}_{k_1}\ldots \delta^{l_{p}}_{k_{p}}
\ee

According to our definition, if $\Gamma^{(n)}$ consists
of two disconnected components $\Gamma^{(n_1)}$ and
$\Gamma^{(n_2)}$, $n = n_1 + n_2$,
then the same is true about $\gamma^{(m)}$,
it also consists of disconnected  $\gamma^{(m_1)}$ and $\gamma^{(m_2)}$
(both can still be disconnected),
$m = m_1 + m_2$, and

\be
\langle \Gamma^{(n)} | g_T | \gamma^{(m)}\rangle \ =
\langle \Gamma^{(n_1)} | g_T | \gamma^{(m_1)}\rangle
\langle \Gamma^{(n_2)} | g_T | \gamma^{(m_2)}\rangle
\label{com}
\ee
It is natural to introduce the product of disconnected graphs
as their unification and then interpret (\ref{com}) as
the group-element property (\ref{comult}) of $g_T$.

From the definition it immediately follows that

\be
\sum_j \langle \Gamma^{(n)}_{i_1\ldots i_n} | g_T |
\gamma_{(m)}^{j_1\ldots j_m}\rangle
\langle \gamma^{(m)}_{j_1\ldots j_m} | g_T |
\tilde\gamma_{(l)}^{k_1\ldots k_l}\rangle   =
\langle \Gamma^{(n)}_{i_1\ldots i_n} | g_T |
\tilde\gamma_{(l)}^{k_1\ldots k_l}\rangle
\label{comultgr}
\ee
for any fixed triple of vertex-subgraphs
$\tilde \gamma \subset \gamma \subset \Gamma$
and given $g_T$.

The basic relation (\ref{mult}) now acquires
the form:

\be
\sum_{all\ \gamma \in {\cal V}\Gamma:
\tilde \gamma \subset \gamma \subset \Gamma}
\langle \Gamma|g_T|\gamma\rangle\langle\gamma|g_{T'}|\tilde\gamma\rangle\
= \ \langle\Gamma| g_{T+T'}|\tilde\gamma\rangle
\ee
for any fixed  $\tilde \gamma \in {\cal V}\Gamma$
and any two $g_T$ and $g_{T'}$.
In more detail, the multiplication relation states:

\be
\sum_m \left(\sum_{all\ \gamma \in {\cal V}\Gamma:
\tilde \gamma \subset \gamma \subset \Gamma}
\left(
\sum_j \langle \Gamma^{(n)}_{i_1\ldots i_n} | g_T |
\gamma_{(m)}^{j_1\ldots j_m}\rangle
\langle \gamma^{(m)}_{j_1\ldots j_m} | g_{T'} |
\tilde\gamma_{(l)}^{k_1\ldots k_l}\rangle  \right)\right) = \nn \\ =
\langle \Gamma^{(n)}_{i_1\ldots i_n} | g_{T+T'} |
\tilde\gamma_{(l)}^{k_1\ldots k_l}\rangle
\label{multgr}
\ee

Let us illustrate the relation (\ref{multgr}) by a couple of examples.

\subsection{Examples}

1)
Let $\tilde\gamma = \emptyset$ and take a double-vertex graph
$\Gamma_{3,4,;2}^{(3)}$ for $\Gamma$.
Then

\be
\langle \Gamma^{(3)}_{i_0;i_1i_2} | g_T | \emptyset\rangle\ =
\sum_{m,n,\tilde m,\tilde n}
T^{(3)}_{i_0mn}G^{m\tilde m}G^{n\tilde n}
T^{(4)}_{i_1i_2\tilde m\tilde n} =
\sum_{mn} T_{i_0}^{mn}T_{i_1i_2 m n}
\ee
For the sake of brevity we omitted
the labels $(3)$ and $(4)$ in coupling constants.

Four different vertex-subgraphs $\gamma$ contribute to the sum in
(\ref{multgr}):

\be
\gamma^{(0)} = \emptyset;\ \
\gamma^{(3)} = \Gamma^{(3)}_{3;0};\ \
\gamma^{(4)} = \Gamma^{(4)}_{4;0} \ \ and\ \
\gamma^{(3)} = \Gamma^{(3)}_{3,4;2}
\ee
The corresponding
\be
\Gamma^{(3)}/\gamma^{(0)} = \Gamma^{(3)};\ \
\Gamma^{(3)}/\gamma^{(3)} = \Gamma^{(4)}_{4;0};\ \
\Gamma^{(3)}/\gamma^{(4)} = \Gamma^{(3)}_{3;0}; \ \
\Gamma^{(3)}/\Gamma^{(3)} = \emptyset
\ee
Eq.(\ref{multgr}) states that

\be
\sum_{m,n} \left(T_{i_0}^{mn}T_{i_1i_2 m n} \cdot 1 +
T_{i_0}^{mn}\cdot {T'}_{i_1i_2 m n} +
{T'}_{i_0}^{mn}\cdot T_{i_1i_2 m n}
+ \right. \nn \\ \left. +
1 \cdot {T'}_{i_0}^{mn}T'_{i_1i_2 m n}\right) =
\sum_{m,n}(T+T')_{i_0}^{mn}(T+T')_{i_1i_2 m n}
\ee
what is indeed true.
Note that in this check it is important that the metric
$G_{ij}$ is the same for all  the three theories
$g_T$, $g_{T'}$ and $g_{T+T'}$.

2)
Let $\Gamma$ be a chain graph
\be
\Gamma^{(2)}_N = C_N
\label{chain}
\ee
with $N$ vertices of valence two, connected by $N-1$ propagators.
Then
\be
\langle \Gamma^{(2)}_N | g_T | \emptyset\rangle_{ij} =
\sum_{\stackrel{i_1,\ldots i_{N-1}}{j_1,\ldots,j_{N-1}}}
T^{(2)}_{ij_1}G^{j_1i_1} T^{(2)}_{i_1j_2}G^{j_2i_2} \ldots
G^{j_{N-1}i_{N-1}}T^{(2)}_{i_{N-1}j}
\ee
In this case $\gamma$ and $\tilde\gamma$ in (\ref{multgr})
can be any collections of disconnected chains of the same type
with the total length of no more than $N$.
If $\tilde\gamma = \emptyset$, there are as many
as $2^N$ possible choices of vertex-subgraphs
$\gamma$ in (\ref{multgr}),
specified by all possible subsets of $N$ crosses in (\ref{chain}).
In particular, there are $\frac{N!}{k!(N-k)!}$ vertex-subgraphs
with $k$ vertices (connected and disconnected),
and in the single-scalar case the
identity (\ref{multgr}) is just the binomial formula:

\be
\frac{1}{G^{N-1}}\sum_{k=0}^N \frac{N!}{k!(N-k)!}
T_{(2)}^k (T'_{(2)})^{N-k}
= \frac{(T_{(2)} + T'_{(2)})^N}{G^{N-1}}
\ee
One can easily restore the indices $i$ and also consider
non-trivial subchains $\tilde\gamma$.

\section{Two Hopf algebras of graphs \label{CK}}
\setcounter{equation}{0}

The universal set-theoretical Hopf algebra defines a product
of two graphs $\Gamma_1$ and $\Gamma_2$ to be a disconnected graph
with components $\Gamma_1$ and $\Gamma_2$,
\be
\Gamma_1\cdot\Gamma_2 = \Gamma_1 \cup \Gamma_2 \ \ \
{\it for}\ \ \ \Gamma_1 \cap \Gamma_2 = \emptyset
\ee
(the role of unity is played by the empty graph $\emptyset$),
and the coproduct
\be
\Delta_{ST}(\Gamma) = \sum_{all\ \gamma \in {\cal V}\Gamma}
\gamma\otimes\Gamma/\gamma
\label{st}
\ee
This Hopf algebra is both commutative and cocommutative,
associative and coassociative. Because of its cocommutativity,
it is not associated with any non-trivial Lie algebra
(the dual algebra ${\it Shift} {\cal M}$,
introduced in (\ref{multgr}), is obviously
commutative: $g_Tg_{T'} = g_{T+T'} = g_{T'}g_T$).
One can define such
a Hopf algebra not only on graphs, but on any set and its subsets
and we call it the ``set-theory'' (ST) Hopf algebra.

For functions on graphs, taking values
in some commutative associative ring ${\cal K}$,
one can define ST multiplication:

\be
(F\odot_{ST} G)(\Gamma) = m((F\otimes G) (\Delta_{ST}(\Gamma)) =
\sum_{\gamma\in {\cal V}\Gamma} F(\gamma)G(\Gamma/\gamma), \nn \\
\ee
(operation $m$ multiplies two components of the tensor product:
$m((F(\gamma_1)\otimes G(\gamma_2)) = F(\gamma_1) G(\gamma_2)$).

Using the specifics of graphs, one can
substitute vertex-subgraphs in (\ref{st})
by box-subgraphs and construct a
non-cocommutative comultiplication \cite{CK}.
First of all, the matrix element of $g_T$ for contracted
graph $[\Gamma/\gamma]$, obtained by contraction of
a box-subgraph $\gamma^{(m)}$ in $\Gamma^{(n)}$, is given by

\be
\langle \left[\Gamma/\gamma\right]_{i_1\ldots i_n}| g_T |
\emptyset \rangle\ = \sum_j
\langle \Gamma^{(n)}_{i_1\ldots i_n}| g_T |
\gamma_{(m)}^{j_1\ldots j_m} \rangle T^{(m)}_{j_1\ldots j_m}
\ee
for connected $\gamma^{(m)}$,

\be
\langle \left[\Gamma/(\gamma_1\cdot\gamma_2)\right]_{i_1\ldots i_n}
| g_T |\emptyset \rangle\ = \nn \\
= \sum_{j,k}
\langle \Gamma^{(n)}_{i_1\ldots i_n}| g_T |
\gamma_{1(m_1)}^{j_1\ldots j_{m_1}} \cdot
\gamma_{2(m_2)}^{k_1\ldots k_{m_2}}
 \rangle T^{(m_1)}_{j_1\ldots j_{m_1}}T^{(m_2)}_{k_1\ldots k_{m_2}}
\ee
for $\gamma^{(m)}$ consisting of two connected parts,
and so on.

The Connes-Kreimer (CK) comultiplication

\be
\Delta_{CK} \Gamma =
\emptyset \otimes \Gamma +
\Gamma \otimes \emptyset +
{\sum_{\gamma\in {\cal B}\Gamma}}
\gamma \otimes\left[\Gamma/\gamma\right], \ \ \
{\it Con}(\Gamma) = 1, \nn \\
\Delta_{CK}(\Gamma_1\cdot\Gamma_2) =
\Delta_{CK}(\Gamma_1)\Delta_{CK}(\Gamma_2)
\label{ckcom}
\ee
and the CK product
\be
(F\odot_{CK} G)(\Gamma) = m((F\otimes G) (\Delta_{CK}(\Gamma)) =
\nn \\ \stackrel{{\it Con}(\Gamma) = 1}{=}\
F(\emptyset)G(\Gamma) + F(\Gamma)G(\emptyset) +
{\sum_{\gamma\in {\cal B} \Gamma}}F(\gamma)G([\Gamma/\gamma])
\label{ckprod}
\ee
are no longer cocommutative. Thus the dual algebra
is the universal
enveloping of non-trivial Lie algebra. This Lie algebra,
${\it diff}_\emptyset {\cal M}$, is
straightforwardly realized by vector fields on the moduli
space ${\cal M}$ of coupling constants.
The universal model (\ref{genthe}) provides a basis in
${\it diff}_\emptyset {\cal M}$, labeled by {\it connected}
graphs: for any connected $\Gamma^{(n)}$ one explicitly defines
$\hat Z_\Gamma \in T{\cal M}$ as

\be
\hat Z_{\Gamma} =
\sum_i \langle \Gamma^{(n)}_{i_1\ldots i_n}
| g_T | \emptyset\rangle
\frac{\partial}{\partial T^{(n)}_{i_1\ldots i_n}} =
\sum_i Z^{(n)}_{i_1\ldots i_n}\{T\}
\frac{\partial}{\partial T^{(n)}_{i_1\ldots i_n}}
\label{vec}
\ee
In what follows we denote by hats the vector fields
and other elements of the universal enveloping
${\cal U}({\it diff}_\emptyset {\cal M})$ to distinguish
them from scalars and other elements of modules (representations)
of ${\it diff}_\emptyset {\cal M}$.
The commutator

\be
\left[ \hat Z_{\Gamma_1}, \hat Z_{\Gamma_2}\right] =
\hat Z_{[\Gamma_1,\Gamma_2]},
\ee
where commutator $[\Gamma_1,\Gamma_2]$ \cite{CK}
is a linear combination of all
graphs $\Gamma$, such that $[\Gamma/\Gamma_1] = \Gamma_2$ --
these enter with the coefficient $+1$,--
or $[\Gamma/\Gamma_2] = \Gamma_1$ -- these enter with $-1$.
(i.e. one blows any valence-$m$ vertex in $\Gamma_2$
by insertion of $\Gamma_1^{(m)}$ and any valence-$n$ vertex
in $\Gamma_1$ by gluing in $\Gamma_2^{(n)}$ and takes an
algebraic sum over such graphs with insertions.)

Disconnected graphs are associated with
higher-order differential operators, e.g.

\be
\hat Z_{\Gamma^{(n_1)}\cdot\Gamma^{(n_2)}} =
\sum_{i,j} \langle \Gamma^{(n_1)}_{i_1\ldots i_{n_1}}
\Gamma^{(n_2)}_{j_1\ldots j_{n_2}}
| g_T | \emptyset \rangle
\frac{\partial^2}{\partial T^{(n_1)}_{i_1\ldots i_{n_1}}
\partial T^{(n_2)}_{j_1\ldots j_{n_2}}} = \nn \\
= \ : \hat Z_{\Gamma^{(n_1)}}  \hat Z_{\Gamma^{(n_2)}}:\
\neq \hat Z_{\Gamma^{(n_1)}}  \hat Z_{\Gamma^{(n_2)}}
\ee
In other words, we associate with disconnected graphs
the {\it normal ordered} products of vector fields,
corresponding to each connected component. This
provides a differential operator of certain order, equal
to the number of connected components.

The vacuum graphs with no external legs define vector
fields in the $T^{(0)}$ direction:

\be
\hat Z_{\Gamma^{(0)}} = Z_{\Gamma^{(0)}}\{T\}\frac{\partial}
{\partial T^{(0)}}
\ee

The matrix elements $\langle \Gamma |g_T| \gamma\rangle$
can be associated either with Beltrami differentials:

\be
\mu_{\Gamma/\gamma} =
\sum_{i,j} \langle \Gamma_{i_1\ldots i_{n}}
| g_T | \gamma^{j_1\ldots j_m} \rangle
dT^{(m)}_{j_1\ldots j_m}
\frac{\partial}{\partial T^{(n)}_{i_1\ldots i_{n}}}
\label{vecGG}
\ee
(for connected $\Gamma$ and $\gamma$),

\be
\mu_{\Gamma/(\gamma_1\cdot\gamma_2)} =
\sum_{i,j,k} \langle \Gamma_{i_1\ldots i_{n}}
| g_T | \gamma_1^{j_1\ldots j_{m_1}} \gamma_2^{k_1\ldots k_{m_2}}\rangle
dT^{(m_1)}_{j_1\ldots j_m} dT^{(m_2)}_{k_1\ldots k_{m_2}}
\frac{\partial}{\partial T^{(n)}_{i_1\ldots i_{n}}}
\ee
(for connected $\Gamma$, $\gamma_1$ and $\gamma_2$) etc;
or with the vector fields

\be
\hat Z_{[\Gamma/\gamma]} =
\sum_{i,j} \langle \Gamma_{i_1\ldots i_{n}}
| g_T | \gamma^{j_1\ldots j_m} \rangle
T^{(m)}_{j_1\ldots j_m}
\frac{\partial}{\partial T^{(n)}_{i_1\ldots i_{n}}}
\label{vec[GG]}
\ee
(for connected $\Gamma$ and $\gamma$).
Note that the only difference between (\ref{vecGG})
and (\ref{vec[GG]}) is in the
letter $d$ in front of $T^{(m)}$,
but it makes a lot of difference.

Operators $g_T$ form a subgroup in abelian group
${\it Shift} {\cal M}$,
which acts transitively on the moduli space ${\cal M}$.
They are complemented by the non-abelian subgroup
${\it Diff}_\emptyset ({\cal M})$ of ${\it Diff}{\cal M}$,
which is generated by the
vector fields $\hat Z_\Gamma$, defined in (\ref{vec}), and is
the stability subgroup of the Gaussian point $T^{(n)} = 0$
(since $\langle \Gamma | g_{T=0} | \emptyset\rangle =
\delta_{\Gamma,\emptyset}$ and all ${Z_\Gamma(T=0)} = 0$).
The moduli space itself can be represented as a homogeneous
factor-space ${\cal M} =
{\it Diff}({\cal M})/{\it Diff}_\emptyset ({\cal M}) =
{\it Shift}({\cal M})/{\it Shift}_\emptyset ({\cal M})$.
The action of ${\it Diff}_\emptyset ({\cal M})$ on non-Gaussian
models is relevant for description of renormalization group
flows in ${\cal M}$.

The Lie algebra ${\it diff}_\emptyset {\cal M}$
of vector fields $\hat Z_\Gamma$ on entire
${\cal M}$ has a variety
of reductions to smaller Lie algebras on subspaces ${\cal M}_{red}
\subset {\cal M}$, i.e. there are Lie algebras associated with
smaller families of models than the universal (\ref{genthe}).
For example, one can consider only interactions of a given
valence, i.e. all $T^{(n)} = 0$ for $n\neq k$, then vector
fields (\ref{vec}) associated with connected graphs with exactly
$k$ external legs form a closed Lie subalgebra.
Alternative reduction is to the tree graphs ($Z_\Gamma = 0$
for any $\Gamma$ with loops).

One can also consider the finite sets of indices $I = \{1,\ldots,N\}$
and accordingly reduced moduli spaces ${\cal M}^{(N)}$, in this
case

\be
Vect ({\cal M}^{(N)}) = \left.Vect({\cal M})\right|_{{\cal M}^{(N)}}
\ee

\section{Inverse operators, projectors and R-Ope\-ra\-tion
\label{invop}}
\setcounter{equation}{0}

Due to (\ref{com}) partition functions are characters of the
graph multiplication:
$\langle\Gamma_1\cdot\Gamma_2|g_T|\emptyset\rangle\ = \
\langle\Gamma_1|g_T|\emptyset\rangle
\langle\Gamma_2|g_T|\emptyset\rangle$,
$\langle\emptyset|g_T\emptyset\rangle\ = 1$.

Characters take values in some commutative associative ring ${\cal K}$
and satisfy:

\be
F(\emptyset) = 1, \nn \\
F(\Gamma_1\cdot\Gamma_2) = F(\Gamma_1)F(\Gamma_2), \nn \\
(F\odot_{ST} G)(\Gamma) = m((F\otimes G) (\Delta_{ST}(\Gamma)) =
\sum_{\gamma\in {\cal V}\Gamma} F(\gamma)G(\Gamma/\gamma), \nn \\
(F\odot_{CK} G)(\Gamma) = m((F\otimes G) (\Delta_{CK}(\Gamma))
= \nn \\ \stackrel{{\it Con}(\Gamma) = 1}{=}\
F(\Gamma) + G(\Gamma) +
{\sum_{\gamma\in {\cal B} \Gamma}}F(\gamma)G([\Gamma/\gamma])
\ee
(operation $m$ multiplies two components of the tensor product:
$m((F(\gamma_1)\otimes G(\gamma_2)) = F(\gamma_1) G(\gamma_2)$).

One can define the inverses (antipodes)
of a character $F$, $F^{-1}_{ST}$,
$F^{-1}_{CK}$, which satisfy
$F^{-1}_{ST}\odot_{ST} F(\Gamma) = \delta_{\Gamma,\emptyset}$,
$F^{-1}_{CK}\odot_{CK} F(\Gamma) = \delta_{\Gamma,\emptyset}$,
by recursive formulas:
\be
F^{-1}_{ST}(\emptyset)= F^{-1}_{CK}(\emptyset) = 1, \nn \\
F^{-1}_{ST}(\Gamma) = -F(\Gamma) -
\sum_{\gamma\in{\cal V}\Gamma;\ \gamma \neq \emptyset,\Gamma}
F^{-1}_{ST}(\gamma)F(\Gamma/\gamma),
\label{F-1} \nn \\
F^{-1}_{CK}(\Gamma) \stackrel{{\it Con}(\Gamma) = 1}{=}\
-F(\Gamma) - {\sum_{\gamma\in{\cal B}\Gamma}}
F^{-1}_{CK}(\gamma)F([\Gamma/\gamma])
\label{FCK-1}
\ee
According to (\ref{multgr}), if
$F(\Gamma) = \langle\Gamma|g_T|\emptyset\rangle$,
then $F^{-1}_{ST}(\Gamma) = \langle\Gamma|g_{-T}|\emptyset\rangle$,
but $F^{-1}_{CK}(\Gamma)$ is given by a more sophisticated expression.
In fact, eq.(\ref{FCK-1}) for $F^{-1}_{CK}(\Gamma)$
is closely associated with Bogolubov's recursive formula, defining
the $R$-operation \cite{Bog}.

Assume that the ring ${\cal K}$ as a linear space can be
decomposed into two components,
${\cal K} = {\cal K}_1 \oplus {\cal K}_2$,
with the help of projectors ${\cal P}_\pm$,
${\cal P}_\pm^2 = {\cal P}_\pm$, ${\cal P}_- = I - {\cal P}_+$:
${\cal K}_\pm = {\cal P}_\pm{\cal K}$.
These projectors can be used to define the
``${\cal P}$-inverse'' (${\cal P}$-antipode)
$PF^{-1}$ of $F(\Gamma)$ \cite{CK}:

\be
{\cal P}_-\left( (PF_\odot^{-1}\odot F)(\Gamma) -
\delta_{\Gamma,\emptyset}
\right) = 0, \nn \\
{\cal P}_+\left(PF_\odot^{-1}(\Gamma)\right) = 0
\label{defPF}
\ee
The second condition makes the definition of
${\cal P}$-antipode unambiguous.
The ordinary inverse $F^{-1}_\odot$ is associated with the
trivial projector ${\cal P}_+ = 0$.
One can easily write down recursive formulae for the
${\cal P}$-antipodes for comultiplications $\odot_{ST}$ and
$\odot_{CK}$ by applying ${\cal P}_-$ to the r.h.s. of
(\ref{F-1}):
\be
PF^{-1}_{ST}(\Gamma) = -{\cal P}_-
\left(F(\Gamma) + \sum_{\gamma\in {\cal V}\Gamma;\
\gamma \neq \emptyset,\Gamma}
PF^{-1}_{ST}(\gamma)F(\Gamma/\gamma)\right),
\label{PST}
\nn \\
PF^{-1}_{CK}(\Gamma) \ \stackrel{{\it Con}(\Gamma) = 1}{=}\
-{\cal P}_- \left(F(\Gamma) +
{\sum_{\gamma\in {\cal B}\Gamma}}
PF^{-1}_{CK}(\gamma)F([\Gamma/\gamma])\right)
\label{PF-1} \label{PCK}
\ee

For projector ${\cal P}_+$, possessing
additional triangular
property w.r.to multiplication in ${\cal K}$, namely

\be
{\cal K}_+\cdot{\cal K}_+ \subset {\cal K}_+,\ \
{\cal K}_-\cdot{\cal K}_- \subset {\cal K}_-
\ee
(i.e. the product of any two elements from ${\cal K}_+$
lies again in ${\cal K}_+$ and similarly for ${\cal K}_-$),
the ST ${\cal P}$-inverse of $F(\Gamma)$ is a character whenever
$F(\Gamma)$ is a character:

\be
PF^{-1}_{ST} (\Gamma_1\cdot\Gamma_2) =
PF^{-1}_{ST} (\Gamma_1) PF^{-1}_{ST} (\Gamma_2)
\ee
if

\be
F(\Gamma_1\cdot\Gamma_2) = F(\Gamma_1) F(\Gamma_2)
\ \ \ \forall\ \Gamma_1, \Gamma_2
\ee
Indeed, assume that this is true for all smaller vertex-subgraphs of
$\Gamma_1$ and $\Gamma_2$. Then

\be
PF^{-1}_{ST} (\Gamma_1\cdot\Gamma_2) =
-{\cal P}_-
\left( F(\Gamma_1\cdot\Gamma_2) +
\phantom{\sum_{\gamma_2\in {\cal B}\Gamma_2}}
\right. \nn \\ \left. +
\sum_{\gamma_1\in {\cal B}\Gamma_1}
\sum_{\gamma_2\in {\cal B}\Gamma_2}
PF^{-1}_{ST} (\gamma_1)PF^{-1}_{ST} (\gamma_2)
F(\Gamma_1/\gamma_1)F(\Gamma_2/\gamma_2) - \right.\nn \\ \left.
\phantom{\sum_{\gamma_2\in {\cal B}\Gamma_2}}  -
F(\Gamma_1)F(\Gamma_2) - PF^{-1}_{ST} (\Gamma_1)PF^{-1}_{ST} (\Gamma_2)
\right)
\label{BB}
\ee
The last two items at the r.h.s. subtract the contributions
from $\gamma_1\cdot\gamma_2 = \emptyset$ and
$\gamma_1\cdot\gamma_2 = \Gamma_1\cdot\Gamma_2$.
The double sum in (\ref{BB}) is equal to the product of two sums,
defining the ST ${\cal P}$-inverses of $\Gamma_1$ and $\Gamma_2$,
which (the sums) are both ${\cal P}$-positive. Due to triangularity
the product is also ${\cal P}$-positive and is eliminated by
${\cal P}_-$. Therefore  (\ref{BB}) states that

\be
PF^{-1}_{ST} (\Gamma_1\cdot\Gamma_2) = {\cal P}_-
\left( PF^{-1}_{ST} (\Gamma_1)PF^{-1}_{ST} (\Gamma_2)
\right) =
 PF^{-1}_{ST} (\Gamma_1)PF^{-1}_{ST} (\Gamma_2)
\ee
The last equality is again implied by triangularity, since
both ${\cal P}$-inverses are ${\cal P}$-negative.

Not every projector is triangular, for example
projection on positive numbers in the ring of reals is not
triangular: the product of two negatives is no longer negative.
A natural triangular projector exists in a ring of Laurent
series $\left\{ A = \sum_{k = -N}^\infty a_kz^k\right\}$:
${\cal P}_+ A = \sum_{k = 0}^\infty a_kz^k$.
The difference ${\cal R} = {\cal P}_+ - {\cal P}_-$
is the $r$-matrix,
widely used in the theory of integrable systems and its applications
(see, for example, \cite{Sem} and also \cite{GLM, MorGLM}).
To get a field-theory model with such ${\cal K}$ one can, for
example, consider the $z$-dependent couplings $T^{(n)} =
\sum_{k=-N}^\infty T^{(n)}_k z^k$ in (\ref{genthe}).
In the study of continuous field theory $z$ rather enters through
regularization of infinite sums (integrals) over indices $i$
in (\ref{genthe}): it can be identified with $d - d_{crit}$
for dimensional regularization \cite{CK} or with $1/M$
for Pauli-Villars regularization etc.

According to (\ref{defPF}), the $R$-operation
\be
F(\Gamma) \longrightarrow RF_\odot(\Gamma) =
(PF_\odot^{-1} \odot F)(\Gamma), \nn \\
{\cal P}_- \left(RF_\odot(\Gamma)\right) = 0,
\label{defROP}
\ee
acting on the
space of functions of graphs, converts any function into a
${\cal P}$-positive (``finite'') one.
Moreover,
since $\odot$-product of characters is again a character,
it converts characters into characters.
The main claim of \cite{CK} is that eq.(\ref{defROP}) for
$\odot_{CK}$
can be considered as group-theory interpretation of
Bogolubov's recursion formula \cite{Bog}.
In sec.\ref{reps} we shall see that more relevant in generic
case is the corepresentation $\hat\odot_{CK}$.

Of course, from algebraic perspective there is nothing
special about continuous theory, divergencies and
dimensional regularization: the only things that matter are
algebraic structures and triangular projectors.

\section{Representations of ${\it diff}_\emptyset {\cal M}$
and ${\cal U}({\it diff}_\emptyset {\cal M})$
in differential operators on ${\cal M}$
\label{reps}}
\setcounter{equation}{0}

Returning to the beginning of sec.\ref{invop},
model (\ref{genthe})
provides $Z_\Gamma\{T\}$ for the role of characters,
if these quantities are considered as functions of $\Gamma$,
and $T$-dependence is not taken into account.
Similar treatment can be given to vacuum diagrams
in generic (\ref{genthe}).
However, it is more adequate to treat $Z_\Gamma\{T\}
= \langle\Gamma|T\rangle$ as describing a transformation between
the functions of $T$ and functions of $\Gamma$ (not obligatory
characters), as suggested in sec.\ref{SPAG} above.

Consider the action of vector fields on
linear modules over ${\cal M}$.
Namely, a vector field

\be
\hat V = \sum_n \left(\sum_{i_1,\ldots, i_n}
V^{(n)}_{i_1\ldots i_n}(T)\frac{\partial}{\partial
T_{i_1\ldots i_n}}\right)
\ee
can  act on a function of $T$-variables $F\{T\}$
with or without free indices:

\be
F\{T\} \rightarrow \hat V\{T\} F\{T\}
\ee

Now we can exploit the power given by the use of the universal
model (\ref{genthe}). It provides a large enough set of functions
on ${\cal M}$ to establish the one-to-one correspondence between
linear combinations of graphs
and invariant functions of coupling constants
(while for smaller models the set of such functions is much smaller:
graphs label different types of contracting indices,
and there should be many enough indices to
distinguish between different contractions). Because of this,
every invariant (i.e. with all indices $i$ contracted with the
help of the metric $G_{ij}$)
function on ${\cal M}$ can be uniquely decomposed into
a sum over graphs
of the basic functions $Z_\Gamma\{T\}$, introduced in sec.\ref{SPAG}
(of course, such expansions survive certain
reductions of ${\cal M}$, but this is a separate story).
Actually, {\bf functions} are decomposed into sums
over vacuum graphs $\Gamma^{(0)}$ without external legs,

\be
F\{T\} = \sum_{\Gamma^{(0)}} F(\Gamma) Z_\Gamma\{T\}
\label{Fdec}
\ee
with $T$-independent coefficients $F_\Gamma$;
{\bf vector fields} -- over connected graphs with any number
of external legs,

\be
\hat V\{T\} = \sum_{n \geq 1} \left(
\sum_{{connected}\ {\Gamma^{(n)}}}
V(\Gamma) \hat Z_\Gamma\{T\}\right);
\label{Vdec}
\ee
the {\bf $k$-differentials} -- over graphs with $k$ connected
components and non-vanishing number of external
legs in each component,
\be
:\hat W_k\{T\}: \ = \nn \\ =
\sum_{n_1,\ldots n_k \geq 1} \left(
\sum_{\stackrel{connected}{\Gamma_1^{(n_1)},\ldots,\Gamma_k^{(n_k)}}}
W(\Gamma_1\cdot\ldots\cdot\Gamma_k)
\ :\hat Z_{\Gamma_1}\{T\}\cdot\ldots\cdot
\hat Z_{\Gamma_k}\{T\}:\right);
\label{Wdec}
\ee
{\bf generic elements} of the universal module (generic differential
operators on ${\cal M}$) -- over all possible graphs (with any number
of connected components and external legs).
In what follows $F\{T\}$ can be arbitrary element of
the universal module. Also, we assume that for a vector field
the coefficients $V(\Gamma)$ are defined for all graphs $\Gamma$,
just $V(\Gamma) = 0$ if $\Gamma$ is not connected
(of course, $V(\Gamma)$ is not a character, characters are
associated with group elements $G_{\hat V} = e^{\hat V}$,
not vector fields themselves).

The result of the action of $\hat V$ on $F$ can also be decomposed
in the basis $Z_\Gamma\{T\}$,

\be
\hat V\{T\} F\{T\} =
\sum_\Gamma (\hat V F)(\Gamma) Z_\Gamma\{T\}
\label{VFdec}
\ee
and
one obtains a relation between the coefficients
$(\hat VF)(\Gamma)$, $F(\Gamma)$ and $V(\Gamma)$: since

\be
\sum_\Gamma (\hat V F)(\Gamma) Z_\Gamma\{T\} =
\sum_{\Gamma'} \sum_{\gamma} V(\gamma) F(\Gamma')
\left(\hat Z_\gamma(T) Z_{\Gamma'}\{T\}\right),
\ee
and

\be
\hat Z_\gamma\{T\} Z_{\Gamma'}\{T\} =
\sum_{\Gamma:\ \Gamma' = [\Gamma/\gamma]} Z_\Gamma\{T\}
\ee
we get a convolution formula

\be
(\hat V F)(\Gamma) =
\sum_{\gamma\in {\cal B}\Gamma}
V(\gamma) F([\Gamma/\gamma]) =
(V\hat\odot_{CK} F)(\Gamma)
\label{coefrel}
\ee
Operation $\hat\odot_{CK}$,

\be
(W\hat\odot_{CK} F)(\Gamma) =
m((W\otimes F)(\hat\Delta_{CK}\Gamma)) = \nn \\
\stackrel{{\it Con}(\Gamma) = 1}{=}\
W(\emptyset)F(\Gamma) + \sum_{\gamma \in {\cal B}\Gamma}
W(\gamma)F([\Gamma/\gamma]),
\label{corepdu}
\ee
is expressed in terms of the
corepresentation of the CK Hopf algebra of graphs,

\be
\hat\Delta_{CK}\Gamma = \emptyset\otimes\Gamma +
\sum_{\gamma \in {\cal B}\Gamma} \gamma\otimes [\Gamma/\gamma],
\ \ \ {\it Con}(\Gamma) = 1,
\label{corep}
\ee
the same way as $\odot_{CK}$, eq.(\ref{ckprod}), is expressed
through the comultiplication

\be
\Delta_{CK}\Gamma = \emptyset\otimes\Gamma + \Gamma\otimes\emptyset +
\sum_{\gamma \in {\cal B}\Gamma} \gamma\otimes [\Gamma/\gamma],
\ \ \ {\it Con}(\Gamma) = 1
\ee

In (\ref{coefrel}) $\hat V$ is a vector field, therefore
$V(\emptyset) = 0$.
The difference between comultiplication $\Delta$ and corepresentation
$\hat\Delta$ is in associativity conditions:

\be
(\Delta \otimes id)\Delta = (id \otimes \Delta)\Delta
\ee
for $\Delta$ and

\be
(\Delta \otimes id)\hat \Delta = (id \otimes \hat \Delta)\hat \Delta
\ee
for $\hat\Delta$.

Repeated application of formula (\ref{coefrel})
defines the action of products
and normal ordered products of vector fields on $F$. For two vectors,
since

\be
\hat Z_{\gamma_1}\left(\hat Z_{\gamma_2} Z_{\Gamma''}\right) =
\sum_{\Gamma:\ [\Gamma/\gamma_1] = \Gamma'}
\left(\sum_{\Gamma':\ [\Gamma'/\gamma_2] = \Gamma''}Z_\Gamma\right)
\ee
and

\be
{:\hat Z_{\gamma_1}\hat Z_{\gamma_2}:}\ Z_{\Gamma'} =
\sum_{\Gamma:\ [\Gamma/(\gamma_1\cdot\gamma_2)] = \Gamma'}Z_\Gamma,
\ee
we have:

\be
(\hat V_1\hat V_2 F)(\Gamma) =
\sum_{\gamma_2\in {\cal B}\Gamma}  V_2(\gamma_2) \left(
\sum_{\gamma_1 \in {\cal B}[\Gamma/\gamma_2]}
V_1(\gamma_1)
F(\left[[\Gamma/\gamma_2]/\gamma_1\right])\right)
\label{coefrelvv}
\ee
and

\be
(:\hat V_1\hat V_2: F)(\Gamma) =
\sum_{\gamma_1\cdot\gamma_2\in {\cal B}\Gamma}
V_1(\gamma_1)V_2(\gamma_2)
F([\Gamma/(\gamma_1\cdot\gamma_2)]) = \nn \\ =
(:V_1V_2:\hat\odot_{CK} F)(\Gamma)
\label{coefrel:vv:}
\ee
Note that in these formulas
$\left[[\Gamma/\gamma_2]/\gamma_1\right] \neq
[\Gamma/(\gamma_1\cdot\gamma_2)]$:
a graph $\gamma_1$ can appear after contraction
$[\Gamma/\gamma_2]$ is made. Relatively simple formulae
in terms of the corepresentation
$\hat\odot_{CK}$ exist only for the normal
ordered elements ${:\hat W:}\ \in {\cal U}({\it diff}_\emptyset
{\cal M})$:\footnote{
Note that the only component of ${\cal U}({\it diff}_\emptyset
{\cal M})$ which has $W(\emptyset) \neq 0$ is a counity,
i.e. $W\{T\} = const$. Non-trivial
functions $\hat f\{T\} \notin {\cal U}({\it diff}_\emptyset
{\cal M})$, and ordinary product of functions is not
expressible in terms of coproduct $\odot_{CK}$. Instead
the scalar
$\hat f\{T\} = \sum_{\Gamma^{(0)}}f(\Gamma) Z_\Gamma\{T\}$ acts on
$F\{T\}$ as

$$
(\hat f F)(\Gamma) =
\sum_{\Gamma_1,\Gamma_2:\ \Gamma = \Gamma_1\cdot \Gamma_2}
f(\Gamma_1)F(\Gamma_2)
$$
In particular, for connected $\Gamma$,
$(\hat f F)(\Gamma) = f(\Gamma) + F(\Gamma)$.
}

\be
({:\hat W:}\ F)(\Gamma) \stackrel{{\it Con}(\Gamma) = 1}{=}\
{:W:}(\emptyset) F(\Gamma) +
\sum_{\gamma\in {\cal B}\Gamma}
{:W:}(\gamma) F([\Gamma/\gamma]) = \nn \\ =
({:W:}\ \hat\odot_{CK} F)(\Gamma)
\label{coefrelgen}
\ee

Of special interest for us are specific elements of
${\cal U}({\it diff}_\emptyset {\cal M})$,
which are the {\it group elements}, \cite{gentau}
and form the diffeomorphism group ${\it Diff}_\emptyset {\cal M}$.

Given a vector field $\hat V = V^\alpha \partial_\alpha$
($\alpha$ is a multiindex, labeling connected graph with
indices or any linear combinations of such graphs),
one can make an element of ${\it Diff}_\emptyset ({\cal M})$
by exponentiation:

\be
G_{\hat V} = e^{\hat V} =
\sum_{n=0}^\infty \frac{\hat V^n}{n!}
\ee
However, it is not normal ordered, and the action of $G_{\hat V}$
on $F(\Gamma)$ is described by sophisticated expression:

\be
(G_{\hat V}F)(\Gamma) =
\sum_{n = 0}^\infty\frac{1}{n!}\left(\sum_{\{\gamma\}_n}
V(\gamma_1)\ldots V(\gamma_n) F([\Gamma/\{\gamma\}_n])\right)
\label{expexp}
\ee
where $\{\gamma\}_n$ denotes a hierarchy of subgraphs
$\gamma_n \in {\cal B}\Gamma$,
$\gamma_{n-1} \in {\cal B}[\Gamma/\gamma_n]$, $\ldots$,
$\gamma_1 \in {\cal B}\left[\left[\ldots \left[
[\Gamma/\gamma_n]/\gamma_{n-1}\right]/\ldots \right]/\gamma_2\right]$.

One can instead expand $G_{\hat V}$ in normal order constituents
with the help of a forest formula:
\be
G_{\hat V} = e^{\hat V} =
\sum_{n=0}^\infty \frac{\hat V^n}{n!} = \nn \\
= 1 + V^\alpha\partial_\alpha +
\frac{1}{2} V^\gamma\partial_\gamma V^\alpha\partial_\alpha
+ \frac{1}{6} V^\gamma\partial_\gamma
V^\beta\partial_\beta V^\alpha\partial_\alpha
 + \ldots = \nn \\  = 1 +
\left(V^\alpha + \frac{1}{2}V^\gamma (\partial_\gamma V^\alpha)
+ \frac{1}{6} V^\gamma(\partial_\gamma
V^\beta)(\partial_\beta V^\alpha) + \right. \nn \\ \left.
+ \frac{1}{6} V^\beta V^\gamma(\partial_\beta\partial_\gamma V^\alpha)
+ \ldots \right)\partial_\alpha + \nn \\ +
\frac{1}{2}\left(V^\alpha +
\frac{1}{2}V^\gamma (\partial_\gamma V^\alpha) + \ldots\right)
\left(V^\beta  + \frac{1}{2}V^\gamma (\partial_\gamma V^\beta)
+ \ldots\right)\partial_\alpha \partial_\beta +  \nn \\ +
\frac{1}{6}
(V^\alpha + \ldots)(V^\beta + \ldots)(V^\gamma + \ldots)
\partial_\alpha \partial_\beta \partial_\gamma +\ldots = \nn \\ =
1 + \sum_{{\cal F}} \frac{1}{{\it Tree}({\cal F})!}
:\prod_{{\cal T}\in {\cal F}}
\frac{\hat V_{{\cal T}}}{\sigma_{\cal T}{\cal T}!} :
\label{forest}
\ee
The forest ${\cal F}$ is an ordered set of rooted trees. Rooted
tree has a single external leg (root), all other external
legs end at the valence-one vertices. ${\it Tree}({\cal F})$
is the number of trees in the forest, and
${\it Vert}({\cal T})$ is the number of vertices in the tree
${\cal T}$.
For every rooted tree
$\sigma_{{\cal T}}$ is the symmetry factor (the order of the
discrete group which interchanges subtrees, leaving the tree
intact), while the tree-factorial \cite{CM} is defined
iteratively: ${\cal T}! = {\it Vert}({\cal T})\prod_a
{\cal T}_a!$, where ${\cal T}_a$ are the root subtrees formed
after the root is cut away.
In every vertex of a tree stands the vector field $\hat V$,
acting on the neighbor vertex downwards (in the direction
to the root), and not further. Then with every tree we
associate a vector field $\hat V_{{\cal T}}$, which contains
the ${\it Vert}({\cal T})$'s power of $\hat V$ and
${\it Link}({\cal T})$ derivatives
(${\it Link}({\cal T})$ is the number of links in the tree).
For example, for the 1-vertex (${\cal T}_1$),
2-vertex (${\cal T}_2$) and 3-vertex/2-branch (${\cal T}_Y$) trees:

\be
V({\cal T}_1) = V^\alpha\partial_\alpha, \nn \\
V({\cal T}_2) = V^{\alpha}(\partial_\alpha V^\beta)\partial_\beta,
\nn \\
V({\cal T}_Y) =  V^{\alpha}V^\beta(\partial_\alpha \partial_\beta
V^\gamma)\partial_\gamma,
\ee
etc. With a forest we associate a differential operator, which
is a normal-ordered product of vector fields $\hat V_{\cal T}$
over the trees (as usual, normal ordering means,
that all derivatives are
written to the right of $V^\alpha$'s, this is a
coordinate-dependent operation, e.g.
$:\hat V^3: = V^{\alpha}V^\beta V^\gamma\partial_\alpha \partial_\beta
\partial_\gamma$).

One can apply (\ref{coefrelgen}) to obtain an alternative
expression to (\ref{expexp}) in terms of the corepresentation
$\hat\odot_{CK}$. Complexity of the formula is now encoded in the
sum over forests.
One can efficiently handle this complexity by the following
trick. Since $G_{\hat V} = e^{\hat V}$ is a diffeomorphism of
${\cal M}$, for any $F\{T\}$  we have:

\be
(G_{\hat V}F)\{T\} = F\{T + \tilde V(T)\}
\label{AA}
\ee
with

\be
\tilde V^{(n)}_{i_1\ldots i_n} = \left(e^{\hat V} - 1\right)
T^{(n)}_{i_1\ldots i_n} =
\left(\sum_{{\cal T}}
\frac{\hat V_{{\cal T}}}{\sigma_{\cal T}{\cal T}!}\right)
T^{(n)}_{i_1\ldots i_n}
\label{AB}
\ee
The first equality in (\ref{AB}) is obtained by substitution of
$T^{(n)}_{i_1\ldots i_n}$ instead of $F\{T\}$ in (\ref{AA}), the second
equality is implied by the forest formula (\ref{forest}), because
a normal product of two or more vector fields annihilates
$T^{(n)}_{i_1\ldots i_n}$. Now introduce a new vector field

\be
\hat{\tilde V}\{T\} = \sum_n\left( \sum_{i_1,\ldots, i_n}
\tilde V^{(n)}_{i_1\ldots i_n} \frac{\partial}{\partial
T^{(n)}_{i_1\ldots i_n}}\right) = \sum_{connected\ \Gamma}
\tilde V(\Gamma)\hat Z_\Gamma\{T\},
\label{AC}
\ee
such that the shift operator

\be
G_{\hat V} = e^{\hat V} = \ {:e^{\hat{\tilde V}}:}
\label{AD}
\ee
Equality (\ref{AD}) is implied by (\ref{AA}) and by
Taylor expansion

\be
F\{T + \tilde V(T)\} =   \ {:e^{\hat{\tilde V}}:}\ F\{T\}
\label{AE}
\ee
Now we can make use of (\ref{coefrelgen}) to obtain a simple
substitute for (\ref{expexp}):

\be
(G_{\hat V}F)(\Gamma)\ \stackrel{{\it Con}(\Gamma) = 1}{=}\
F(\Gamma) + \sum_{\gamma \in {\cal B}\Gamma} G_{\hat V}(\gamma)
F([\Gamma/\gamma]) = \nn \\ =
\sum_{n=0}^\infty
\left(\sum_{\stackrel{non-intersecting}{\gamma_1,\ldots,\gamma_n \in
{\cal B}\Gamma}} \tilde V(\gamma_1) \ldots \tilde V(\gamma_n)
F([\Gamma/(\gamma_1\cdot\ldots\cdot\gamma_n)]\right)
\label{Bogol}
\ee

\section{Bogolubov's recursion and renormalized Lagrangian}
\setcounter{equation}{0}

One can apply diffeomorphisms in moduli space to ``improve''
partition functions. This is important if one wants to eliminate
undesired dependence on one or another
parameter of the theory, like ultraviolet cut-off in continuous
local field models. Basically, one needs to project the
entire moduli space ${\cal M}$ onto certain subspace ${\cal M}_{ren}$
of ``renormalized models''. The problem is that
pa\-ra\-me\-ter-de\-pen\-dence
arises in partition functions, and arbitrary elimination of
unwanted parameters from particular correlators
can break the relation to
Lagrangian formalism and moduli space. It is exactly the problem,
which is resolved by Bogolubov's $R$-operation \cite{Bog},
and which can be most straightforwardly described in terms
of diffeomorphisms of ${\cal M}$.

The $R$-operation can be formulated as follows:

Given (triangular) projectors ${\cal P}_\pm$ in the ring ${\cal K}$
(where matrix elements and partition functions are taking values)
and a function $F\{T\}$ (with or without free indices, i.e. any
element of the universal module over ${\cal M}$),
one finds a specific diffeomorphism $G_{\hat P} \in
{\it Diff}_\emptyset{\cal M}$ which makes $F\{T\}$ ${\cal P}$-positive:

\be
{\cal P}_-\left(F\{T + \tilde P(T)\}\right) = 0,
\label{POP0}
\ee
\be
F\{T+\tilde P(T)\} = e^{\hat P} F\{T\} = (G_{\hat P} F)\{T\}
\label{POP1}
\ee
To define such diffeomorphism unambiguously, one imposes
additional constraint on $\hat P\{T\}$, for example,

\be
{\cal P}_+ \left( \tilde P^{(n)}_{i_1\ldots i_n}\right) = 0
\ \ \forall\ n;\ i_1,\ldots,i_n
\label{POP2}
\ee
(see eq.(\ref{POP2'}) below for a more adequate constraint).
Some constraint of this type is needed to distinguish between
``renormalizations'', needed to eliminate ${\cal P}$-negative
contributions to the correlation functions from arbitrary
diffeomorphisms of ${\cal M}$, which can map ${\cal P}$-positive
models into other ${\cal P}$-positive ones.

Eqs.(\ref{POP0}-\ref{POP2}) define the Bogolubov's
$R$-operation for any projector ${\cal P}_+$. One can apply the
machinery of the previous sections to rewrite (\ref{POP1}) either
in terms  of Gauss-Birkhoff decomposition of the shift operator,

\be
g_{T+\tilde P(T)} = g_T g_{\tilde P(T)}
\label{deco}
\ee
(decomposing ${\cal P}$-positive renormalized model
into the bare one and ${\cal P}$-negative counterterm model),
or in terms of CK algebra of functions on graphs.
In the last case one can use any of the three representations
(\ref{expexp}), (\ref{forest}) or (\ref{Bogol}).
The most convenient is the third choice, and it is exactly
the one providing the Bogolubov's recursion formula.
Eq.(\ref{Bogol}) can indeed be rewritten in the form of
a recurrent relation for $\tilde P(\Gamma)$, expressing it
through $\tilde P(\gamma)$ for smaller box-subgraphs $\gamma$
(with less vertices), provided $F(\Gamma)$ does not vanish on
elementary vertices $[\Gamma/\Gamma]$. Indeed, one can extract from the
r.h.s. of (\ref{Bogol}) two items: one with $n = 0$
and another with $n=1$ and $\gamma = \Gamma$.
Then we obtain:\footnote{
In notation of \cite{CK} $G_{\hat P}(\Gamma) = C(\Gamma)$.
}

\be
\tilde P(\Gamma) F([\Gamma/\Gamma])\ \stackrel{{\it Con}(\Gamma) = 1}{=}\
-{\cal P}_-\left(F(\Gamma) +  \sum_{\gamma \ \in\ {\cal B}\Gamma;\
\gamma\neq\Gamma}G_{\hat P}(\gamma)F([\Gamma/\gamma])\right)
= \nn \\ =
-{\cal P}_-\left(F(\Gamma) + \sum_{n=1}
\left(\sum_{{\gamma_1,\ldots,\gamma_n \in
{\cal B}'\Gamma}} \tilde P(\gamma_1) \ldots \tilde P(\gamma_n)
F\left([\Gamma/(\gamma_1\cdot\ldots\cdot\gamma_n)]\right)
\right)\right)
\label{Bogol2}
\ee
Here $\gamma \in {\cal B}'\Gamma$ means that the
sum goes over {\it non-intersecting}
box-subgraphs $\gamma_1,\ldots,\gamma_n \neq \Gamma$.
We also assumed that
(\ref{POP2}) is in fact substituted by a more sophisticated constraint

\be
{\cal P}_+\left( \tilde P(\Gamma) F([\Gamma/\Gamma]) \right) = 0
\label{POP2'}
\ee
Then we can omit ${\cal P}_-$ acting on the l.h.s. of (\ref{Bogol2}).
Eq.(\ref{Bogol2}) provides a recursion formula for $\tilde P(\Gamma)$
if $F([\Gamma/\Gamma])
\neq 0$ whenever the r.h.s. of (\ref{Bogol2}) is non-vanishing.
In fact, this is a necessary requirement for renormalizability of
the theory (of a particular reduction of the universal model
(\ref{genthe})): all the elementary vertices $[\Gamma/\Gamma]$ should
be included into the bare Lagrangian, if they have the structure
which can be generated in perturbation theory with
${\cal P}$-negative coefficients
(in traditional language of quantum field theory:
if there are divergent diagrams with a given
number of external legs and external-momenta dependence, an elementary
vertex with such valence and momentum dependence
should be included into the bare Lagrangian).

Recursive formula (\ref{Bogol2}) has a formal solution in
terms of its own forest formula, involving decorated rooted trees.
For connected graph $\Gamma$
consider a sequence of embedded box-subgraphs,
complementary to $\{\gamma\}_n$ in (\ref{expexp}),
$\{\{\gamma\}\}_n: \ \gamma_0 = \Gamma,\ \gamma_1 \in
{\cal B}\Gamma,\ \gamma_2 \in {\cal B}\gamma_1 \subset
{\cal B}\Gamma, \ldots, \gamma_n \in {\cal B}\gamma_{n-1} \subset
\ldots \subset {\cal B}\gamma_1 \subset {\cal B}\Gamma$.
It corresponds to a collection of non-intersecting boxes,
which can now (in variance with the set, used in the
definition of particular box-subgraph) lie one inside another.
Such collection allows one to build
a decorated rooted tree ${\cal T}$ \cite{CK}.
If $\Gamma$ is disconnected, there will be trees, associated
with every connected component.
With lower site of each box one associates a vertex of the
tree, two vertices are connected by a link if one of the
corresponding boxes lies immediately inside another (i.e.
there are no boxes in between the two). The root link ends
at a vertex, associated with $\gamma_0 = \Gamma$.
According to this construction, every vertex of the tree
is associated with connected box-subgraph $\hat\gamma_k
\subset \gamma_k$ ($\gamma_k$ need not be connected),
and there is exactly one link, going downwards (towards the
root, i.e. associated with the neighbor bigger box) and
connecting $\hat\gamma_k$ to some $\hat\gamma_{k-1}$,
and unrestricted number of links, going upwards and connecting
$\hat\gamma_k$ to some collection $\hat\gamma_{k+1}^1,\ldots,
\hat\gamma_{k+1}^{s(k)} \subset \gamma_{k+1}$.
The solution to (\ref{Bogol2}) associates with every vertex
$\hat\gamma_k$ an operator
$\left(F([\hat\gamma_k/\hat\gamma_k])\right)^{-1}
(-{\cal P}_-) F([\hat\gamma_k/(\hat\gamma_{k+1}^1\cdot\ldots\cdot
\hat\gamma_{k+1}^{s(k)})])$, where projector ${\cal P}_-$
acts upwards along the branches of the tree. The root vertex
$\hat\gamma_0$ (i.e. a connected component of $\Gamma$)
contributes just $ F([\hat\gamma_0/(\hat\gamma_{1}^1\cdot\ldots\cdot
\hat\gamma_{1}^{s(0)})])$. In these terms the result of
$R$-operation can be written as follows \cite{Bog}:

\be
(G_{\hat P}F)(\Gamma) = \nn \\ =
\sum_{{\cal F}_\Gamma}
\prod_{{\cal T}\in{\cal F}_\Gamma}
\left( \prod_{\stackrel{vertices}{of\ {\cal T}}}^{\longrightarrow}
\frac{1}{F([\hat\gamma_k/\hat\gamma_k])}(-{\cal P}_-)
F([\hat\gamma_k/(\hat\gamma_{k+1}^1\cdot\ldots\cdot
\hat\gamma_{k+1}^{s(k)})]
\right)
\ee
Arrow over the product sign means that the product
is ordered along the branches.

Importance of Bogolubov's recursion in the space of function $F\{T\}$
is that it converts partition functions ($\tau$-functions)
into partition functions,
while arbitrary subtraction procedure, like the naive
$F_\Gamma\{T\} \rightarrow {\cal P}_-(F_\Gamma\{T\})$, does not have
this property: it may
not be represented as an action of ${\it Diff} {\cal M}$
and no operator $g_{T+\tilde P(T)}$ results from such a subtraction.

It  deserves noting that
$\left(F\{T+\tilde P(T)\}\right)_\Gamma \neq
F_\Gamma\{T+\tilde P(T)\}$.
For example, for the simplest chain graph $C_1$
(one valence-two vertex)

\be
Z_{C_1}^{ij}\{T+\tilde P(T)\} = T_{(2)}^{ij} +
\tilde P_{(2)}^{ij}\{T\} = \nn \\ =
Z_{C_1}^{ij}\{T\} + \tilde P_{C_1}^{ij}\{T\} +
\tilde P_{C_2}^{ij}\{T\} +
\ldots,
\ee
while

\be
\left(Z^{ij}\{T+\tilde P(T)\}\right)_{C_1} =
Z_{C_1}^{ij}\{T\} + \tilde P_{C_1}^{ij}\{T\}
\ee
Because of this difference one sometime says that renormalization
of Lagrangian does not make contribution of each individual graph
${\cal P}$-positive (in the sense that sometime
${\cal P}_-\left(Z_\Gamma\{T+\tilde P(T)\}\right) \neq 0$),
while $R$-operation does  (in the sense that always
${\cal P}_-\left(Z\{T+\tilde P(T)\}\right)_\Gamma = 0$).
However, as we just explained, if interpreted properly, renormalization
of Lagrangian and $R$-operation are just the same.

From here on -- if one wants to continue -- one needs to
split the universal model (\ref{genthe}) into smaller
universality classes, which differ by the choice and properties of
the sets $I$ (where indices $i$ in (\ref{genthe}) take values),
especially by the ways the possibly-divergent sums over indices
(e.g. integrals over momenta) are regularized (it still makes sense
to keep the full set of coupling constants $T^{(n)}_{i_1\ldots i_n}$).
The most interesting projectors exploit particular properties of
particular $I$'s.
They can act non-trivially on the basic functions $Z_\Gamma\{T\}$,
not only on the coefficient functions $F(\Gamma)$
(this actually happens in the case of regularized continuous
field models, at least in the naive approach).
For particular projectors the counter-terms $\tilde P(\Gamma)$
can vanish for certain classes of graphs (for
divergency-eliminating projectors in renormalizable field models
contributing are only graphs with loops and restricted
number of external legs).
Given $I$ and ${\cal P}_\pm$, one can say that
the $R$-operation (\ref{POP1}) provides a full set of
${\cal P}$-positive functions on ${\cal M}(I)$: a linear basis
is provided by the set of $Z_\Gamma\{T + \tilde P(T)\}$
(generically this space is smaller than the one with the basis
$Z_\Gamma\{T\}$).

\section{Conclusion}

We described the relation between the algebraic structures,
introdu\-ced by A.Con\-nes and D.Krei\-mer, and the generic bilinear
relations (Hirota equations) for effective
actions in quantum field theory. We discussed
two groups acting on
the moduli space ${\cal M}$ of theories: one, essentially
commutative ${\it Shift} {\cal M}$, acting transitively on ${\cal M}$
and responsible for bilinear relations;
another, the non-commutative stability subgroup of the Gaussian
point ${\it Diff}_\emptyset {\cal M}$ in the diffeomorphism group
${\it Diff}{\cal M}$,
is associated with Lie
algebra of vector fields on ${\cal M}$, it is related to
the CK Hopf algebra of graphs,
to Bogolubov's $R$-operation and to renormalization group flows.
Bogolubov's $R$-operation is defined in terms of projector
operators and can be expressed as renormalization
of the action ($T$-dependent shift of the coupling constants $T$).
This study provides a
long awaited support to the idea of hidden integrability of
non-perturbative quantum phenomena from
the field of conventional field theory (Feynman diagram
technique).

It also opens a way for the study of analogous phenomena
in perturbative string theory, where graphs are substituted
by open Riemann surfaces and CK Hopf algebra has interesting
generalizations (an infinitesimal deformation in that direction
is to the Hopf algebra of fat graphs, associated with the
universal matrix model (\ref{genmathe}).

The old belief that the moduli space ${\cal M}$ of theories
and diffeomorphism group ${\it Diff}{\cal M}$ are indeed very similar
to conventional simple moduli spaces, studied in mathematics
and elementary string theory, gains a new support from  the
observations in earlier papers of D.Krei\-mer \cite{DK}.
However, this subject is beyond the scope of the present paper.

\section{Acknowledgments}

We are indebted for discussions to A.Gorsky, A.Levin,
A.Losev and V.Novikov.

Our work is partly supported by RFBR grants
98-01-00328 (A.G.), 98-02-16575 (A.M.),
by INTAS grant 97-0103 (K.S.),
by Scientific Schools Grant 00-15-96557 (A.G.)
and by the Russian President's Grant 00-15-99296 (A.M.).

\end{document}